\documentclass{article}
\usepackage{graphicx} 

\title{Modelling the elliptical instability of magnetic skyrmions}
\author{Bruno Barton-Singer}

\usepackage{amsmath,amsfonts,amsthm,amscd,amssymb,slashed,graphicx,caption,subcaption,a4wide,exscale,relsize,setspace,lineno}
\usepackage{macros}

\begin{document}

\maketitle

\begin{center}
African Institute of Mathematical Sciences (AIMS) Mbour, Senegal \\ bruno.s.b.singer@aims-senegal.org
\end{center}

\section*{Abstract}
{
Two recently developed methods of modelling chiral magnetic soliton elliptical instability are applied in two novel scenarios, the tilted ferromagnetic phase of chiral magnets dominated by easy-plane anisotropy and the general case of the chiral magnet with tilted applied field and arbitrary uniaxial anisotropy. In the former case, the analytical predictions are found to exactly match previous numerical results. In the latter case, instability of isolated chiral skyrmions has not yet been studied, although the predictions correspond interestingly to previous numerical investigation of the phase diagram.
}

\section{Introduction\label{sec_introduction}}

Magnetic skyrmions are localised textures found in chiral magnets \cite{Bogdanov89,Muhlbauer09,Yu10} that in various ways are particle-like, such as having an associated conserved charge, moving in response to external forces \cite{Papanicolaou91}, exerting forces on each other \cite{Bogdanov95,Foster19} and even forming crystals that undergo phase transitions \cite{Huang20}. An essential feature of magnetic skyrmions is that they are topologically non-trivial, meaning that starting with a skyrmion, no continuous transformation of the magnetisation field respecting the boundary conditions can give us the uniform ferromagnetic state. Skyrmions can thus not be destroyed, provided the physical dynamics obeys these constraints. This robustness offers the possibility of applications in data storage \cite{Fert13}. Recently, a large variety of similar textures in chiral magnets have been theoretically investigated \cite{Foster19,Rybakov19,Kuchkin20oct}, which we call in general chiral magnetic solitons. Many but not all of these solitons are similarly topologically protected, and offer promising alternatives to the conventional magnetic skyrmion \cite{Gobel11}.


However, one of the remarkable features of chiral magnetic solitons is that despite this topological nature they can have a variety of instability modes. These include elliptical instabilities where the solitons elongate indefinitely into domain walls, as well as collapse instabilities where they shrink to a point in finite time. Thus the solitons can in fact be destroyed, either by violating the boundary condition or continuity. It is thus a fundamental question in the study of chiral magnetic solitons to understand how these instabilities arise and at what material parameters.

These instabilities were first studied for axisymmetric chiral skyrmions in \cite{Bogdanov94b}, taking into account the interaction of the skyrmion with the stray field which in this paper we neglect. The study of radial, but not elliptical, stability of general axisymmetric solitons was treated in \cite{Bogdanov99}. These papers employed the direct method of calculating the Hessian of the energy at the specific solution considered, using certain ans\"{a}tze to bound the instability or numerics to approximate it directly. More recently, the stability of axisymmetric solutions has been proven analytically in the presence of Zeeman interaction \cite{Melcher18} and a combination of Zeeman interaction and easy-axis anisotropy \cite{Li18}, in either case only in the limit where the coefficients of these terms in the energy functional are sufficiently large.

The larger class of chiral magnetic solitons are for the most part not axisymmetric, and yet these solitons all experience elliptical instability at similar material parameters \cite{Kuchkin20oct}. The question naturally arises of whether there is a unifying soliton-independent model that describes the elliptical instability. In \cite{Kuchkin20oct} the authors found that the elliptical instability could be modelled as arising from a negative energy per unit length of the isolated $2\pi$-domain wall in the system, and this simple model gives good numerical agreement for a general chiral magnet with both normally applied field and uniaxial anisotropy, provided the anisotropy is not so large that it changes the background uniform state, a case we discuss below.

At the same time, we can consider the elliptical instability of skyrmions and antiskyrmions in tilted applied field \cite{Kuchkin20feb}. We find that this first method is not sufficient to model the elliptical instability, and a novel method is necessary: in \cite{BartonSinger22} the authors used the linearised Euler-Lagrange equations to calculate the decay lengthscale of the tail of the soliton, and calculated for what parameters this decay lengthscale diverges, corresponding to a change in the linear stability of the uniform state. The resulting neutral stability line closely approximates numerical observations in a complementary region of the phase diagram to the first method.

In this paper, we build on these two methods for predicting the elliptical instability and extend them to other cases of interest, namely the tilted ferromagnetic phase arising in chiral magnets with sufficiently large easy-plane anisotropy, and the tilted applied field case considered above, but allowing also for the presence of easy-axis or easy-plane anisotropy. 

We call the former case the symmetry-breaking phase for clarity, as in contrast to the case of tilted applied field the chiral magnet is itself axisymmetric but minimisers of the appropriate energy functional spontaneously break this symmetry. In this phase, skyrmions are also called bimerons \cite{Kharkov17}. Isolated skyrmion stability in the symmetry-breaking phase was numerically modelled in \cite{Leonov17}.

The latter case of tilted applied field and uniaxial anisotropy represents the generic situation when a tilted magnetic field is applied to a sample, as uniaxial anisotropy is a material property and not in general equal to zero. This regime is numerically investigated in \cite{Leonov17b}, although the question of isolated skyrmion stability is not directly considered.

The paper is structured as follows: we start by reviewing the chiral magnet model in Sec. \ref{sec_chiral_magnet}. We allow for a more general potential than is normally considered, meaning that even here some new calculations are necessary to determine the uniform state of the system. We then review the two methods discussed above, the domain wall energy method in Sec. \ref{sec_dw_method} and the diverging lengthscale method in \ref{sec_m=0_method}. In Sec. \ref{sec_results} we then present the predictions given by these two methods for the boundary in the phase diagram where the onset of elliptical instability is seen, and compare to numerical results where they are available. Sec. \ref{sec_symmetry-breaking_results} discusses the results for solitons in the symmetry-breaking phase of the chiral magnet, and Sec. \ref{sec_tilted_field_anis_results} the results for solitons in the chiral magnet with tilted applied field and non-zero anisotropy.

\section{The chiral magnet model\label{sec_chiral_magnet}}

In this paper we consider the two-dimensional chiral magnet energy functional:

\bee
E(\bn)= \int\left( \frac{1}{2}\partial_i\bn\cdot\partial_i\bn +\bD_i\cdot(\bn\times\partial_i\bn) + V(\bn) \right)d^2x.
\label{chiral_magnet_energy}
\eee

The quantities $\bD_1,\bD_2$ are the Dzyaloshinskii-Moriya Interaction (DMI) \cite{Dzyaloshinskii58,Moriya60} vectors representing the preferred orientation and rate of twisting of the magnetisation along the $x_1$ and $x_2$ axes respectively. In general this gives us six independent material parameters, but it is natural to consider the case where the system is symmetric under rotation around an axis normal to the plane: if we denote rotation by $\theta$ anticlockwise around the axis $\be$ by $R(\theta,\be)$, this symmetry is represented by the transformation

\bee
\bn(\vx) \mapsto \bn'(\vx) = R(\theta,\be_3)\bn(R(-\theta,\be_3)\vx),
\label{rotation_transformation}
\eee

and the most general DMI satisfying this symmetry, which we call axisymmetric DMI, is given by $\bD_i=-k R(\beta,\be_3)\be_i$, where $k$ and $\beta$ are two free parameters describing the strength of interaction and the preferred orientation of rotation respectively: $\beta=0$ corresponds to Bloch-type DMI, $\beta=\frac{\pi}{2}$ corresponds to N\'{e}el-type DMI, and intermediate $\beta$ corresponds to a material with a combination of the two. We restrict ourselves to this two-parameter family of DMI in this paper as it is the most commonly considered, although the methods used here can be applied to the general case. We neglect the boundary term considered in \cite{Schroers19}, as it has no bearing on these results. 

DMI of this kind happens to also be invariant under a reflection-like symmetry. If we use $P(\be)$ to denote reflection in a plane perpendicular to $\be$, then the following transformation:
\bee
\bn(\vx)\mapsto P\left(R\left(\beta-\frac{\pi}{2},\be_3\right)\be\right)\bn(P(\be)\vx), \quad \be\perp\be_3.  
\label{reflection-like_symmetry}
\eee
leaves the DMI invariant. So for Bloch-type DMI, the spin and the spatial co-ordinate are reflected simultaneously in two perpendicular planes that are also perpendicular to the plane of the magnet. For N\'{e}el-type DMI the two planes are parallel, and for a generic combination of DMI the planes are at an intermediate angle. For the rest of this paper we will set $\beta=0$ without loss of generality, as models with different values of $\beta$ are related by a rotation of $\bn$.


Although the methods used in this paper are general, we will specify the physically relevant potential
\bee
V(\bn)=h_z(1-\be_h\cdot\bn)+h_a(1-n_3^2).
\label{general_potential}
\eee

This is made up of two commonly considered contributions. The first is the Zeeman interaction, coming from an applied magnetic field along $\be_h$ of strength $h_z$. We can define the direction of $\be_h$ by the angles $\theta_h$ and $\phi_h$:

\bee
\be_h=\begin{pmatrix}
\sin\theta_h\cos\phi_h\\
\sin\theta_h\sin\phi_h\\
\cos\theta_h
\end{pmatrix}
\label{ehdefinition}
\eee

The second commonly-considered contribution is the perpendicular anisotropy coming from the crystal structure itself, characterised by the parameter $h_a$. Positive $h_a$ corresponds to easy-axis anisotropy, favouring $\bn=\pm\be_3$, while negative $h_a$ corresponds to easy-plane anisotropy, favouring $\bn\perp\be_3$. As $\phi_h$ can be set to zero without loss of generality, our general potential is parametrized by ($h_z$, $h_a$, $\theta_h$), with $h_z>0$.

The first question that must be answered for a general potential is the location of the absolute minimum or minima. This gives the orientation of the uniform ferromagnetic state, which would be an absolute energy minimum of the energy functional in the absence of DMI. In the presence of DMI, a helical or skyrmion lattice state can be lower energy than the uniform state \cite{Bak80,Han10}. However,
we are interested in the stability of isolated solitons within the uniform ferromagnetic state, so it is an essential starting point for both methods discussed in Sec. \ref{sec_methods}.

If there is a unique absolute minimum, we denote this $\bn_0$. If there is a set of absolute minima, we choose one without loss of generality. Using angular co-ordinates on the sphere, we describe $\bn_0$ by the angles $\Theta_0$, $\Phi_0$:

\bee
\bn_0=\begin{pmatrix}
\sin\Theta_0\cos\Phi_0\\
\sin\Theta_0\sin\Phi_0\\
\cos\Theta_0.
\end{pmatrix}
\label{n0definition}
\eee

The simplest case is that of the potential \eqref{general_potential} with $\theta_h=0$:

\bee
V(\bn)=h_z(1-n_3) + h_a(1-n_3^2)
\label{axisymmetric_potential}
\eee
this potential is commonly considered as it is experimentally simplest to apply a field perpendicular to the sample. It is symmetric under $\bn\to R(\theta,\be_3)\bn$, so we call it the axisymmetric potential. When $h_z+2h_a \geq 0$, $\bn_0=\be_3$ so the vacuum retains the symmetry, $\Theta_0=0$ and we call this the symmetry-retaining phase. In this phase the isolated magnetic skyrmion retains this symmetry \cite{Bogdanov89}, while other more recently discovered magnetic solitons do not \cite{Kuchkin20oct}. 

When $h_z+2h_a<0$, the minima of the potential lie on a circle with $\Phi_0$ arbitrary and

\bee
\Theta_0=\arccos\left(-\frac{h_z}{2h_a}\right).
\label{symmetry_breaking_vac_angle}
\eee

In this case there is spontaneous symmetry-breaking as $\bn=\be_3$ is a local maximum, and we call this the symmetry-breaking phase.

\begin{figure}
	\centering
	\includegraphics[width=0.6\linewidth]{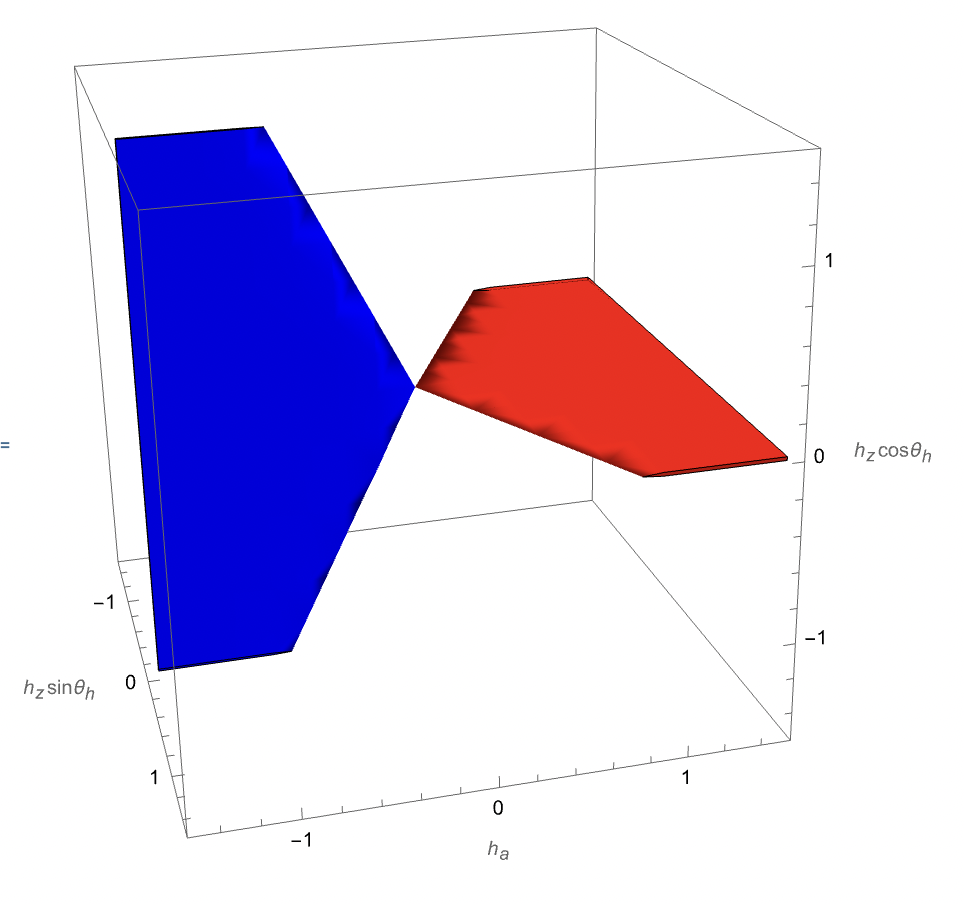}
	\caption{Phase diagram of the vacua of the general potential with easy-axis ($h_a>0$) or easy-plane ($h_a<0$) anisotropy and Zeeman interaction of strength $h_z$ tilted at $\theta_h$ to the normal of the plane, leading to an in-plane component of the magnetisation $h_z\sin\theta_h$ and an out-of-plane component $h_z\cos\theta_h$. The ambient space around the blue and red surfaces represents potentials with a unique minimum, which varies continuously as a function of $(h_z,h_a,\theta_h)$. Crossing through either surface leads to the vacuum changing discontinuously. The blue surface represents potentials with a $U(1)$ symmetry that is spontaneously broken, resulting in a tilted ferromagnetic phase. The red surface represents potentials with a $\mathbb{Z}_2$ symmetry $n_3\to-n_3$ that is again spontaneously broken.}
	\label{fig:3D_phase_diagram}
\end{figure}

When we consider non-zero $\theta_h$, the dependence of $\Theta_0$ and $\Phi_0$ on ($h_z$, $h_a$,$\theta_h$) is in general a complicated function, which is discussed in detail in Appendix \ref{appendix_minima}. The resulting phase transitions, where $\Theta_0$ is discontinuous as a function of the material parameters, are plotted in Fig. \ref{fig:3D_phase_diagram}.

Qualitatively, $\Theta_0$ is approximately equal to $\theta_h$ when $h_z$ dominates over $h_a$. Meanwhile for a larger positive $h_a$ there can be a first-order phase transition as $\theta_h$ crosses $\frac{\pi}{2}$, and for a larger negative $h_a$ there can be a first-order transition as $\theta_h$ crosses 0. These two first-order transitions form two-dimensional surfaces within the phase diagram, each of which corresponds to a phase where there are multiple minima of the potential and thus spontaneous symmetry-breaking. The phase transition at $\theta_h=0$ is the $U(1)$ symmetry-breaking phase described above. The phase transition at $\theta_h=\frac{\pi}{2}$ is a $\mathbb{Z}_2$ symmetry-breaking phase that is less studied but has some interesting features, such as skyrmion-antiskyrmion exchange symmetry \cite{Kuchkin20feb}. 

In the presence of DMI, the boundary of the symmetry-breaking phase is special in that there exists an exact axisymmetric skyrmion solution, and the point $h_z=-2h_a=k^2$, where $k$ is the DMI strength, is doubly special in that an infinite number of solutions can be found, including the axisymmetric skyrmion and skyrmionium solutions but also many more \cite{Schroers18}. This critical coupling corresponds to two points  $(h_z,h_a,\theta_h)=(k^2,-0.5k^2,0/\pi)$ on the edge of the blue surface in Fig. \ref{fig:3D_phase_diagram}. It is natural to ask if the boundary of the other symmetry-breaking phase is special in a similar way.

\section{Methods\label{sec_methods}}
\subsection{The zero-energy domain wall method\label{sec_dw_method}}


If we take a 1D cross-section through the axisymmetric skyrmion in any direction, or a specific cross-section through the antiskyrmion, we see something that is qualitatively like a single $2\pi$-domain wall: it has the same symmetries and twists in the correct direction to lower the DMI, but it does not have the exact profile of a minimiser of the 1D chiral magnet because of the influence of the rest of the soliton. However, as the soliton extends in the direction perpendicular to the cross-section, we might expect that the cross-sectional profile starts to approximate a 1D domain wall more closely. We see this explicitly at the `critical coupling' point of the chiral magnet phase diagram \cite{Schroers18}: here there is no energy barrier to the extension of a skyrmion or antiskyrmion into a domain wall and as the (anti-)skyrmion extends, its cross-sectional profile approaches the profile of the $2\pi$-domain wall exactly. The energy per unit length of this domain wall is zero, so at this critical point we can think of domain walls as extensible objects with no tension, and this lack of tension is then what allows various other solitons to extend to infinity.

This inspires us to consider approximating the cross-sectional profile of any elliptically unstable soliton by the appropriate $2\pi$-domain wall. As the soliton extends, this approximation should get better as the influence of the ends of the soliton reduces. If the energy density of the wall is negative, no matter how small, according to this argument a sufficiently extended soliton has negative energy per unit length and it is energetically favourable for it to extend further. Meanwhile if it is positive per unit length, elliptical instability to infinity should not be allowed.

Note that if an isolated domain wall has negative energy per unit length, then the uniform ferromagnetic background cannot be the ground state: a single isolated domain wall will have lower energy, and two isolated domain walls placed sufficiently far apart lower energy than that, etc. So this calculation in fact reveals the location of the boundary between a ferromagnetic and a helical groundstate. This is interesting in itself, as in the situations we consider in this paper such a boundary has only been found numerically, by minimizing within the class of periodic 1D solutions and comparing the energy to the uniform state. Here we recreate these boundaries with explicit and exact formulae. It is also important to note that even if the helical groundstate is lower energy than the uniform ground state, the energy barrier to transition may be too high and the uniform ground state is metastable. On the other hand, the energy barrier to the elongation of a single soliton will be much lower.


To find the energy-minimizing domain wall in a given system, we use the principle of symmetric criticality \cite{Palais79}: if we take the most general field configuration satisfying a certain subset of the symmetries of the energy, substitute it into the energy, and look for the Euler-Lagrange equations that arise for this restricted set of field configurations, then the solution to these equations will also be a stationary point of the full Euler-Lagrange equations. In particular, asking for a solution that satisfies the reflection-like symmetry \eqref{reflection-like_symmetry} combined with translation symmetry perpendicular to the spatial plane of reflection constrains us to the standard domain wall ansatz, namely a Bloch wall for Bloch-type DMI and a N\'{e}el wall for N\'{e}el-type DMI.

In the standard case of the axisymmetric potential \eqref{axisymmetric_potential}, we are free to choose the orientation of this reflection, leading to a domain wall oriented along an arbitrary axis. However in the cases considered in this paper, the rotation symmetry \eqref{rotation_transformation} is either explicitly or spontaneously broken, leading to only a single reflection symmetry, which we can take without loss of generality to be $n_2\mapsto -n_2$, $x_1\mapsto -x_1$, and thus the translation symmetry is $x_1\mapsto x_1+a$. We are then constrained to the domain wall ansatz

\bee
\bn(x) =\begin{pmatrix}
\sin(\Theta_0+f(x_2))\\
0\\
\cos(\Theta_0+f(x_2))
\end{pmatrix}.\qquad 
\lim_{x_2\to-\infty}f(x_2)=0,\quad\lim_{x_2\to\infty}f(x_2)=\pm 2\pi
\label{dw_ansatz}
\eee

When this ansatz is substituted into the chiral magnet energy functional, the DMI becomes a boundary term:

\bee
E(f) = \int \left(\int \left(\frac{1}{2} (\partial_2 f)^2 + V(\Theta_0+f)\right)dx_2  \mp 2\pi k\right)dx_1
\eee
and we see that $f(x_2)\to +2\pi$ gives the lower energy so we continue by choosing that boundary condition. 

In general, this integral will be infinite but we are interested in calculating the difference between the domain wall energy and the uniform state per unit length, which we denote $\bar{E}$:

\bee
\bar{E}(f) = \int \left(\frac{1}{2} (\partial_2 f)^2 + V(\Theta_0+f)-V(\Theta_0)\right)dx_2  - 2\pi k
\eee

We can then complete the square in the normal way, bearing in mind we are looking for solutions with everywhere-positive gradient:

\begin{align}
\bar{E}(f) = \int \frac{1}{2} \left(\partial_2 f - \sqrt{2(V(\Theta_0+f)-V(\Theta_0))}\right)^2 dx_2
 +\int_0^{2\pi} \sqrt{2(V(\Theta_0+f)-V(\Theta_0))}df- 2\pi k.
\end{align}

This tells us that solutions of the Bogomol'nyi equation $\partial_2 f = \sqrt{2(V(\Theta_0+f)-V(\Theta_0))}$ with boundary condition $\lim_{x_2\to-\infty}f=0$, $\lim_{x_2\to\infty}f=2\pi$ minimise $\bar{E}$. Such solutions exist if and only if $ V(\Theta_0+f)-V(\Theta_0)>0$ for all $f\neq 0\text{ mod }2\pi$. Supposing this is satisfied, then the energy per unit length of the minimising domain wall solution relative to the uniform state is

\begin{align}
\bar{E}_{\text{min}} =\int_0^{2\pi} \sqrt{2(V(\Theta_0+f)-V(\Theta_0))}df- 2\pi k.
\label{dw_method_general}
\end{align}

$\bar{E}_{\text{min}}$ can be considered as a function of material parameters, and the curve $\bar{E}_{\text{min}}=0$ in the phase diagram is then the prediction of this method for the onset of magnetic soliton elliptical instability.


In \cite{Kuchkin20oct}, this boundary was calculated in the simple case of the symmetry-retaining phase of the axisymmetric potential. Good agreement was found between the numerically observed elliptical instability of various chiral magnetic solitons and this analytical boundary in general. The largest disagreement was found for the magnetic skyrmion, which continues to be stable for small negative energy domain wall energy per unit length. This could in principle be explained by energy barriers to the initial elongation that are soliton-dependent. In \cite{BartonSinger22}, the calculation was extended to solitons in a chiral magnet with tilted applied field and zero anisotropy. In this case as field tilt increases the prediction becomes in general an `under-prediction' of instability, there being a progressively larger region of the phase diagram where domain wall energy per unit length is positive but nevertheless elliptical instability is observed.


Extending this calculation to non-zero anisotropy is straightforward, as the symmetries of the system remain the same, and the only difficulty is that the energy of the domain wall is most simply expressed in terms of the angle $\Theta_0$, which is itself a complicated function of the underlying material parameters $\theta_h$, $h_z$, $h_a$ as discussed in Appendix \ref{appendix_minima}. In the symmetry-breaking phase, the condition that $V(\Theta_0+f)-V(\Theta_0)$ has only one absolute minimum is violated and thus there is no appropriate domain wall solution satisfying the symmetries required, which we discuss further in Sec. \ref{sec_symmetry-breaking_results}.

\subsection{The diverging lengthscale method\label{sec_m=0_method}}

In \cite{BartonSinger22}, the domain-wall method was found to only be a good approximation to elliptical instability for magnetic field tilted at a small angle to the normal. However, a second method was found by considering the profile of the soliton far from its centre, which we call the soliton tail. As $\bn$ approaches $\bn_0$ far from the soliton, the Euler-Lagrange equations are approximately linearised and the decay lengthscale of the soliton tail can be explicitly found in terms of the material parameters of the model by solving these linearised equations, allowing for an arbitrary inner boundary condition representing the soliton core where the linearisation breaks down. Due to the influence of the DMI, material parameters can be found where this decay lengthscale diverges.

Again, this coincides with a phase transition: the divergence of the decay lengthscale can equally be seen as an eigenvalue of the Hessian at the uniform state crossing from positive to negative. We can thus also see this calculation as a calculation of linear stability of the uniform state. Unlike the other method, where even if the true groundstate has changed the uniform state may still be metastable, in this case as we cross the boundary the uniform state becomes unstable and the background state must change.


The question remains of why this phase transition should correspond to soliton instability, and the elliptical instability in particular. In principle, a soliton could remain stable even as the background within which it is embedded changes. However, for the cases we consider here, we will find that the decay lengthscale becomes highly dependent on the direction away from the soliton core, with the decay along one axis remaining unchanged and the decay lengthscale diverging to infinity first along the perpendicular axis.  Thus in these cases, more so than in the case where the method was introduced, it is conceptually clear why this is a model of the elliptical instability.

Concretely, we calculate the linearised Euler-Lagrange equations by substituting $\bn=\bn_0+\psi_1\bbe_1 + \psi_2\bbe_2 +O(\psi_i^2)$ into the Euler-Lagrange equations, where $(\bbe_1,\bbe_2)$ are an orthonormal frame oriented so that $\bbe_1\times\bbe_2=\bn_0$. For general potential, the resulting equation only depends on $a_i=\bD_i\cdot\bn_0$,  and the squared magnon masses $\mu_1^2$, $\mu_2^2$, defined as the eigenvalues of the mass matrix $\frac{\partial^2 V(\psi_1,\psi_2)}{\partial \psi_i\partial \psi_j}$:

\bee
\begin{split}
(-\partial_i\partial_i +\mu_1^2)\psi_1  +2 a_i\partial_i\psi_2  =0  \\
(-\partial_i\partial_i +\mu_2^2)\psi_2  -2 a_i\partial_i \psi_1  =0
\label{general_linearised_euler_lagrange}.
\end{split}
\eee

The vector $\va=(a_1,a_2)$ can be interpreted as the part of the DMI that favours twisting around the axis $\bn_0$, with $\va$ as the preferred direction in the plane along which twisting happens, while $\mu_i$ represents the strength of the restoring force to set $\psi_i$ to zero. In general, $\va$ depends on $\Theta_0,\Phi_0$. In the case of axisymmetric DMI we consider here, $\Phi_0$ is always arbitrary and $\va$ is only a function of $\Theta_0$. For the classes of potential that considered here, we can take $\bbe_1=(\cos\Theta_0,0,-\sin\Theta_0)$, $\bbe_2=(0,1,0)$ and so $\mu_1^2$, $\mu_2^2$ can be written more explicitly as

\bee
\mu_1^2=\frac{\partial^2V}{\partial\Theta^2}
\Bigg\rvert_{\Theta_0,\Phi_0},\qquad \mu_2^2=\frac{1}{\sin^2\Theta_0}\frac{\partial^2V}{\partial\Phi^2}\Bigg\rvert_{\Theta_0,\Phi_0}.
\label{masses_defn}
\eee

In \cite{BartonSinger22}, these equations were derived for the special case $\mu_1=\mu_2$, when the potential is `asymptotically isotropic'. This gives the equations a $U(1)$ symmetry on rotation around $\bn_0$, which allows us to write the equations as a single complex partial differential equation, with a known solution in terms of modified Bessel functions. 

In the case of $\mu_1 \neq \mu_2$, there is probably no explicit solution to \eqref{general_linearised_euler_lagrange}. However, we are only concerned with the decay lengthscale, so we only solves this equation to a further degree of approximation by neglecting terms of order $\frac{\psi}{r}$. In the $\mu_1=\mu_2$ case, this is equivalent to approximating the modified Bessel function solutions by their $e^{-mr}r^{-\frac{1}{2}}$ leading term. Our results for the profile of soliton tails are thus only valid at a large distance from the soliton core relative to the results for the asymptotically isotropic case. This procedure is carried out in Appendix \ref{appendix_2d_decaylengthscale}, but in the main section we can consider a toy model of the system that gives  the same result with less involved calculations.

Our toy model is to consider the long-range tails of a one-dimensional localised structure such as a domain wall, oriented at an arbitrary angle, rather than the tails of a two-dimensionally localised structure like the skyrmion. Explicitly we consider a configuration that only varies along one direction, parametrised by $s$, at an angle $\phi_{\text{DW}}$ to $\va$. We get the coupled equations 

\begin{align}
(-\partial_s^2 +\mu_1^2)\psi_1(s)  +2 \lvert\va\rvert\cos\left(\phi_{\text{DW}}\right)\partial_s \psi_2(s)  =0  \\
 (-\partial_s^2 +\mu_2^2)\psi_2(s)  -2 \lvert\va\rvert\cos\left(\phi_{\text{DW}}\right)\partial_s \psi_1(s)  =0
 \label{1d_el}.
 \end{align}

We then look at the asymptotic behaviour of these equations as $s\to \pm \infty$. For $\cos\phi_{\text{DW}}\neq 0$ and $\mu_1\neq\mu_2$, $\mu_i>0$, we can see that the only non-trivial solutions will have both $\psi_1$ and $\psi_2$ with leading term proportional to $e^{\lambda s}$ for an unspecified (possibly complex) $\lambda$. We then find the matrix equation 
\bee
\begin{pmatrix}
    -\lambda^2 + \mu_1^2 && 2\lvert\va\rvert\lambda\cos\phi_{\text{DW}}  \\
    -2\lvert\va\rvert\lambda\cos\phi_{\text{DW}}  && \lambda^2+\mu_2^2
\end{pmatrix}
\begin{pmatrix}
    \psi_1\\
    \psi_2
\end{pmatrix}=
\begin{pmatrix}
    0\\
    0
\end{pmatrix}
\eee

Setting the determinant of the resulting matrix to zero, we have a quartic equation for $\lambda$ solved by
\bee
\lambda = \frac{\pm 1}{\sqrt{2}}\sqrt{\mu_1^2+\mu_2^2  -4\lvert \va\rvert^2\cos^2\phi_{\text{DW}} \pm \sqrt{(\mu_1^2+\mu_2^2 -4\lvert\va\rvert^2\cos^2\phi_{\text{DW}})^2 - 4\mu_1^2\mu_2^2}}.
\label{lambda_soln}
\eee

The value $\lambda$ can have an imaginary part, which will give an oscillating component to $\psi_1$, $\psi_2$, but its real part must be negative to satisfy our boundary conditions, meaning we take the negative sign on the outside above. We then find that the decay lengthscale is finite for a wall oriented at $\phi_{\text{DW}}$ to $\va$ if

\bee
\mu_1+\mu_2 > 2 \lvert \va\rvert \lvert\cos\phi_{\text{DW}}\rvert.
\label{divergence_condition_dw_1d}
\eee

If $\mu_2=0$, the equation for $\lambda$ simplifies to
\bee
\lambda=0,\pm\sqrt{\mu_1^2-4\lvert\va\rvert^2\cos^2\phi_{\text{DW}}}.
\eee

This appearance of zero as a root of $\lambda$ reflects that there is a zero mode around the uniform state, namely rotation of $\bn_0$ around $\be_3$. From this point of view, where we are really calculating the linear stability of the uniform state, it is not relevant as it remains a zero-mode independent of any change of material parameters. Any change in stability must come from the other roots of $\lambda$. Therefore we continue by discarding the $\lambda=0$ root and considering at what material parameters the other roots of $\lambda $ cross zero.

However, this zero of $\lambda$ is problematic from the point of view that we are calculating the leading behaviour of these domain wall tails (or in Appendix \ref{appendix_2d_decaylengthscale}, the 2D soliton tails). It leaves open the possibility that the real tails of a soliton in this phase have polynomial dependence, and to calculate this dependence we would need to re-include quadratic terms in the Euler-Lagrange equation. This is beyond the scope of this paper, so we just leave open the possibility that the explicit decay lengthscales we calculate for the $\mu_2=0$ case may not be the leading-order behaviour of the tails. Nevertheless, their divergence has physical meaning. Turning to the other roots of $\lambda$, we see that the condition for finite decay length (or linearly stable uniform state) is included within the general condition \eqref{divergence_condition_dw_1d} by na\"{i}vely setting $\mu_2=0$.

Returning to the general case, when $\phi_{\text{DW}}=\frac{\pi}{2}$ the two equations decouple and can have two different decay lengthscales given by $\mu_1$ and $\mu_2$. We see these two lengthscales in the limit of $\phi_{\text{DW}}\to\frac{\pi}{2}$ in the general equation \eqref{lambda_soln}. This means that a one-dimensional structure oriented perpendicular to $\va$ will not see the DMI at all asymptotically, and thus have the same decay lengthscale regardless of the size of $\va$. Meanwhile, the wall oriented parallel to $\va$ will be the first to diverge in lengthscale, at $\mu_1+\mu_2 =2\lvert\va\rvert$. This thus seems a good toy model of elliptical instability, which this argument predicts will always occur along $\va$ when $\mu_1+\mu_2 \leq 2\lvert\va\rvert$.

In Appendix \ref{appendix_2d_decaylengthscale} we repeat this procedure in two dimensions, but like in the $\mu_1=\mu_2$ case we must first Fourier transform in the azimuthal direction, and the single derivative terms become mixing terms between different Fourier components. The end result is that the decay lengthscale varies azimuthally to match the 1D decay lengthscale in any direction, i.e. if we take polar co-ordinates $(r,\phi)$ with $\va$ along the x-axis without loss of generality,

\bee
\psi_{1,2}(r,\phi)\to \exp\left(\frac{- 1}{\sqrt{2}}\sqrt{\mu_1^2+\mu_2^2  -4\lvert\va\rvert^2\cos^2\phi\pm \sqrt{(\mu_1^2+\mu_2^2 -4\lvert\va\rvert^2\cos^2\phi)^2 - 4\mu_1^2\mu_2^2}}r\right)
\label{2d_decay}
\eee
as $r\to\infty$ for given $\phi$.

This means that
\bee
\mu_1+\mu_2 >2\lvert \va\rvert
\eee
is the condition for overall exponential decay, and as we approach this boundary the lengthscale of decay parallel to $\va$ will be the first to diverge.

Up to this point, this argument applies for any DMI, including low-symmetry DMI. In the axisymmetric DMI case that we consider in this paper, the stability condition becomes
\bee
\mu_1+\mu_2 >2k\sin\Theta_0,
\label{diverging_lengthscale_result_mu1mu2theta0}
\eee

and so to find the predicted instability in a given model it remains to substitute the values of $\mu_1$, $\mu_2$ and $\Theta_0$, which can all be found from the potential $V(\bn)$.

\section{Results\label{sec_results}}
\subsection{Skyrmions in the symmetry-breaking phase of the chiral magnet\label{sec_symmetry-breaking_results}}

In the symmetry-breaking phase, the domain wall method is not a useful way of modelling soliton instabillity. Within the domain wall  ansatz \eqref{dw_ansatz}, $\bn$ is constrained to lie along a line of longitude of the sphere, which crosses through the set of minima at two points, meaning that the condition leading to equation \eqref{dw_method_general} is violated. Therefore there is no $2\pi$-domain wall solution, but a `long' and a `short' wall connecting the two vacua, which repel each other so that they cannot co-exist statically. If we imagine a soliton extending along $x_1$ such that its cross-section looks like a $2\pi$-domain wall, then as the influence of the ends of the soliton reduces this domain wall itself disintegrates, as the short and long wall repel each other, giving an expansion in the $x_2$ direction as well. Thus there is no regime where we expect the cross-section of the skyrmion to look like a static minimiser of the energy.

\begin{figure}
\centering
\includegraphics[width=\linewidth]{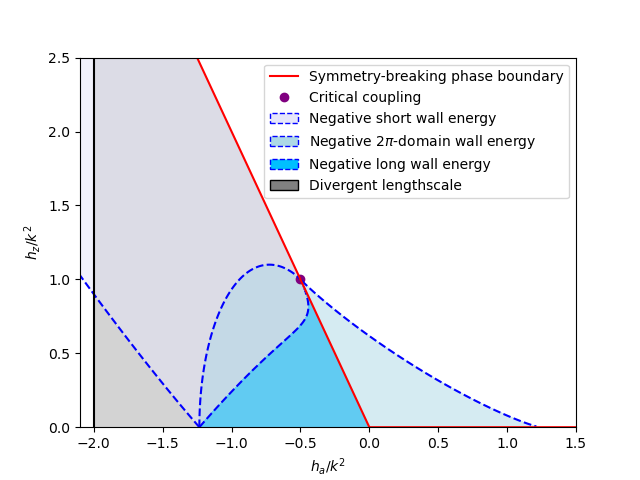}
\caption{Comparison of the different analytical methods of estimating soliton elliptical instability within the axisymmetric chiral magnet, including both symmetry-retaining and symmetry-breaking phase. Within the symmetry-breaking phase there are two possible domain walls, short and long, whose energy per unit length is calculated in \cite{Ross20}. This gives rise to three boundaries as the sign of different domain wall energies change. Also within the symmetry-breaking phase, our diverging lengthscale method predicts elliptical instability for $h_a\geq-2k^2$, and this is borne out by numerical simulation \cite{Leonov17}. The critical coupling point \cite{Schroers18} is marked with a purple dot.}
\label{fig:elliptical_instability_hzs_has}
\end{figure}

Nevertheless, we can ask when the energy per unit length of the short and long domain walls changes sign, as well as the sum of the two. This was calculated explicitly in \cite{Ross20}. When both are positive, their combined energy is certainly positive, and instability by extension of the domain wall might seem unlikely (however, we see below that this assumption is incorrect). Meanwhile, when the sum of the energies per unit length of the two walls becomes negative, then there will be a helical solution that is lower energy than the uniform groundstate, as discussed above. The different contours are plotted in Fig. \ref{fig:elliptical_instability_hzs_has}. We see that when the sum of the two domain wall energies is negative, instability is indeed observed, but this region of the phase diagram is only a subregion of the region of instability predicted by the diverging lengthscale method, which exactly matches numerical results. Meanwhile, if we consider the larger region where just the `short' wall of the system has negative energy, it predicts instability where there is none in one region of the phase diagram and predicts stability where there is instability in another.

The diverging lengthscale method, meanwhile, is surprisingly accurate. In the symmetry-breaking phase $\Theta_0$ is given by \eqref{symmetry_breaking_vac_angle} and according to \eqref{masses_defn} the magnon masses are

\bee
\mu_1^2=-2h_a\left(1-\left(\frac{h_z}{2h_a}\right)^2\right), \; \mu_2^2=0.
\eee

If we substitute this into \eqref{diverging_lengthscale_result_mu1mu2theta0} the equation simplifies to predict elliptical instability when 
\bee
h_a\geq -2k^2.
\eee

This is exactly the numerical result described in \cite{Leonov17}.  The authors in that paper described the instability as elliptical, and here we see concretely the connection to the elliptical instability of skyrmions in tilted applied field, and by extension normal axisymmetric skyrmions. We plot this boundary together with the boundaries calculated from the domain wall method in Fig. \ref{fig:elliptical_instability_hzs_has}.

\subsection{Skyrmions and antiskyrmions in the chiral magnet with tilted applied field and anisotropy\label{sec_tilted_field_anis_results}}

Finally we consider the general case of tilted magnetic field and non-zero anisotropy. Like for the zero anisotropy case, both methods are applicable here.

To find the energy of the isolated domain wall relative to the uniform state we can substitute the given potential into the formula \eqref{dw_method_general} and find

\bee
\begin{split}
\bar{E}_{\text{min}}=4\sqrt{h} + \sqrt{\frac{2}{h_a}}\Bigg((h-h_a)\cos\Theta_0\log\left(\frac{\sqrt{h}+\sqrt{2h_a}\cos\Theta_0}{\sqrt{h}-\sqrt{2h_a}\cos\Theta_0}\right)  \\
+(h+h_a)\sin\Theta_0\arctan\left(\sqrt{\frac{2h_a}{h}}\sin\Theta_0\right)\Bigg) -2\pi k,
\end{split}
\eee

where $h=h_z\sqrt{1-\frac{h_a^2}{h_z^2}\sin^2(2\Theta_0)} + 2h_a\cos(2\Theta_0)$, and $\Theta_0$ is a function of $(h_z,h_a,\theta_h)$, expressed implicitly in the equation \eqref{general_potential_el}. 

Meanwhile, for the diverging lengthscale method we can find $\mu_1$ and $\mu_2$  in terms of $\Theta_0$ from \eqref{masses_defn}:

\bee
\begin{split}
\mu_1^2&= h_z \cos(\Theta_0 -\theta_h) + 2h_a\cos(2\Theta_0)\\\
\mu_2^2 &= h_z\frac{\sin\theta_h}{\sin\Theta_0}
\label{mu1mu2_tiltplusanis}
\end{split}
\eee

We substitute these into \eqref{diverging_lengthscale_result_mu1mu2theta0} to find the predicted elliptical instability. As discussed in Appendix \ref{appendix_minima}, $\Theta_0(\theta_h)$ is a very complicated function while $\theta_h(\Theta_0)$ is simple. So for the purpose of plotting and analysis, we can view $\theta_h$, $\mu_1$, $\mu_2$ and the energy of the isolated domain wall as functions of $\Theta_0$.

\begin{figure}
\centering
\includegraphics[width=0.75\linewidth]{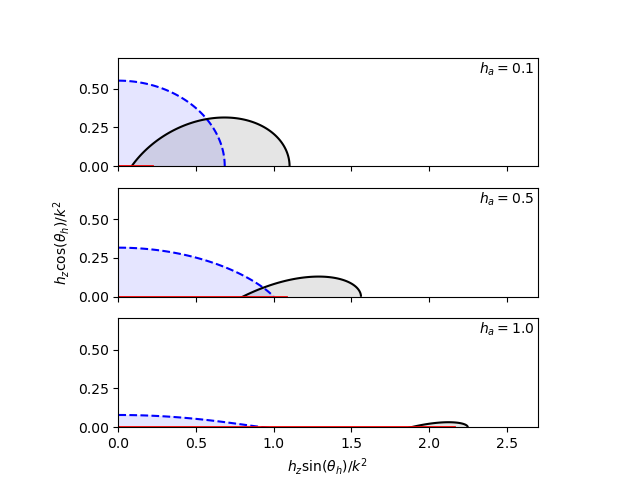}
\caption{Skyrmion region of elliptical instability according to the domain wall method (blue, dashed border) compared to skyrmion region of instability according to divergence of decay lengthscale (grey, solid border), and first-order transition of the vacuum (red) in the case of easy-axis anisotropy.}
\label{fig:elliptical_instability_tiltedfield_pos_ha}
\end{figure}

\begin{figure}
\centering
\includegraphics[width=0.75\linewidth]{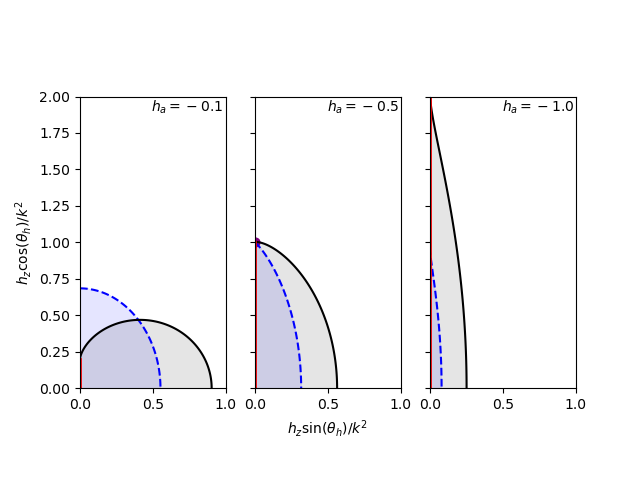}
\caption{Skyrmion region of elliptical instability according to the domain wall method (blue, dashed border) compared to skyrmion region of instability according to divergence of decay lengthscale (grey, solid border), and first-order transition of the vacuum (red) in the case of easy-plane anisotropy. The critical coupling discussed in \cite{Schroers18} is seen in the middle plot at the intersection of the three curves, marked by a purple dot.}
\label{fig:elliptical_instability_tiltedfield_neg_ha}
\end{figure}

The boundaries from the two methods are plotted for easy-axis anisotropy in Fig. \ref{fig:elliptical_instability_tiltedfield_pos_ha} and for easy-plane anisotropy in Fig. \ref{fig:elliptical_instability_tiltedfield_neg_ha}. In both cases we also include the presence of the first-order transitions of the vacuum, discussed in Appendix \ref{appendix_minima}. In the latter case, the critical coupling point \cite{Schroers18} appears at the intersection of the two instabilities.

Unlike in the previous cases, we do not directly have a numerical observation of isolated soliton instability with which to compare these plots. However, there is an interesting congruence with \cite{Leonov17b}. In that paper, the authors look for the lowest-energy phase of the chiral magnet in this setting, numerically measuring the relative energies of the helical, skyrmion lattice and uniform states. The region of instability predicted due to negative domain wall energy, shaded blue in these figures, seems to correspond to the helical phase, which is expected as we have commented above. More surprisingly, the region of instability due to divergent decay lengthscales, shaded grey in these figures, seems to correspond to the skyrmion lattice phase. This brings into sharper focus the question raised in \cite{BartonSinger22}, of whether we are seeing one universal elliptical instability or two different ones. 

The advantage of an explicity expression for our different boundary is that we can go beyond comparison with numerics and make general observations for arbitrary values of $(h_z,h_a,\theta_h)$. For easy-axis anisotropy, the most striking feature is the complete separation of the two instability regions. As the easy-axis anisotropy increases, the region of instability predicted from divergent decay lengthscale moves to larger values of $h_z$ and $\theta_h\simeq \frac{\pi}{2}$, disappearing entirely at $h_z=4k^2$, $h_a=2k^2$. The region of instability predicted from negative domain wall energy shrinks towards $h_z=0$, disappearing at $h_a=8k^2$.

Meanwhile with increasing easy-plane anisotropy, the regions of instability predicted by the two methods overlap more and more, the decay lengthscale method eventually completely taking over as it does in the symmetry-breaking case. The critical coupling point is seen to be the unique point where the boundaries predicted by the two methods intersect with each other and also with the boundary of a first-order transition.

In this case eventually the instability predicted by negative domain wall energy disappears entirely, followed by the instability predicted by diverging lengthscale. We can see that this will occur at $h_a=-\frac{\pi^2}{8}k^2$ and $h_a=-2k^2$ respectively.

In general, we can say that these methods predict that easy-axis anisotropy stabilises systems with a larger out-of-plane component of applied magnetisation, while easy-plane anisotropy stabilises systems with a larger in-plane component of applied magnetisation, while ultimately both have a stabilising effect when they are sufficiently strong.

\section{Conclusion}

In this paper we review two heuristic methods of modelling the elliptical instability of solitons in chiral magnets, which were respectively introduced in \cite{Kuchkin20oct} and \cite{BartonSinger22}. We apply them to two new cases, the symmetry-breaking phase of axisymmetric chiral magnets and the general case of the chiral magnet with tilted applied field and uniaxial anisotropy. In the former case we find exact agreement with numerical results for the onset of elliptical instability, where the question was studied in \cite{Leonov17}. In the latter, we give a prediction that is yet to be numerically tested.

We also give analytical arguments that our two methods for approximating the onset of elliptical instability are also exact predictions of phase transitions where the uniform state is no longer the lowest in energy. In the case of domain wall energy, this is the point at which the helical spin state becomes lower energy than the uniform state, although the uniform state may remain metastable. For the diverging lengthscale method, we only know that the uniform state has become unstable, and it is not clear what the new groundstate is, although numerical results suggest that we are finding the transition to the skyrmion lattice phase \cite{Leonov17b}. Further study is needed to understand if such an instability mode always corresponds to a lattice phase and why.

Unlike previous approaches, the methods presented here are not soliton-specific, but instead find boundaries in the phase diagram which, as they are approached, we would expect to lead to the elliptical instability of all non-trivial textures in the system. The results are thus simple enough to be calculated explicitly, and at the same time widely applicable. 

Most directly, these methods can be extended to systems with non-axisymmetric DMI and other natural potentials. Non-axisymmetric DMI is relevant not only in materials with low crystal symmetry, but also in considering travelling soliton solutions: for example in antiferromagnets, travelling soliton solutions are related to static solutions by a Lorentz-like boost \cite{Kosevich90}, and thus travelling solitons effectively experience an anisotropic DMI. In experiments, as antiferromagnetic skyrmions are pushed to higher velocities they increasingly distort \cite{Jin16}. Simulations indicate a maximum possible antiferromagnetic skyrmion velocity \cite{Komineas20} at which the skyrmion extends to infinity, and the method outlined in this paper could shed light on this observation.

These methods can also be applied to solitons in three dimensions, such as skyrmion tubes or Hopfions, leading to an analytical understanding of the phase diagram of three-dimensional chiral magnets. It is particularly useful in liquid crystal or ferroelectric systems, where there are a large number of parameters describing the system: a single analytical calculation could include all of these, giving the range of material parameters for which solitons could be stable and thus guiding future numerical investigations.

\section*{Acknowledgments}

I thank Paul Sutcliffe, Des Johnston and Calum Ross for helpful comments, and Stavros Komineas for pointing out the possibility of polynomial tails to solitons in the tilted ferromagnetic phase. I also thank the University of Crete for hosting me for an extended research visit during the finalisation of this paper.

\appendix

\section{Minima of the general chiral magnet potential\label{appendix_minima}}

For the general case of tilted field and non-zero anisotropy, finding the stationary points of the potential means solving the equation

\bee
\frac{\partial V}{\partial\Theta}\bigg\rvert_{\Theta_0}=\sin\left(\Theta_0-\theta_h\right)+\frac{h_a}{h_z}\sin(2\Theta_0)= 0
\label{general_potential_el}
\eee

for $\Theta_0$, and comparing the value of the potential in the case of multiple stationary points to find the true minimum. This can be done explicitly, but the resulting functional expression for $\Theta_0(\theta_h,h_z,h_a)$ is unwieldy. Below we calculate the important features of the phase diagram without using this function. 

Firstly, we see that the minimum $\Theta_0$ is only a function of $\theta_h$ and $\rho=\frac{h_a}{h_z}$. The system can be understood by considering three limits: $\rho\to 0$, $\rho \to +\infty$ and $\rho\to-\infty$. In the first case, $\Theta_0=\theta_h + O\left(\rho\right)$, so as $\theta_h$ varies $\Theta_0$ varies continuously. There is only one minimum of the potential so perturbations to the potential do not signicantly change the minima. In the second case when $h_z=0$, $h_a>0$, the potential has two degenerate minima of the potential at $\Theta_0=0, \pi$. Perturbing the potential with non-zero $h_z$ breaks this degeneracy, with $\Theta_0=0$ for $\theta_h<\frac{\pi}{2}$ and $\Theta_0=\pi$ for $\theta_h>\frac{\pi}{2}$. Therefore, as we vary $\theta_h$ across $\frac{\pi}{2}$ we see a first-order transition. In the third case when $h_z=0$, $h_a<0$, the potential has a circle of degenerate minima as parametrized by $\Phi_0$. Given that we are considering applied field at $\phi_h=0$ without loss of generality, we see that positive $h_z$ breaks this degeneracy, picking $\Theta_0=\text{sgn}(\theta_h)\arccos\left(\frac{-h_z}{2h_a}\right)$. Therefore there is a first-order transition as $\theta_h$ crosses $0$.

To check that there are no other transitions, we consider the function
\bee
\rho(\Theta_0)=\frac{\sin(\theta_h-\Theta_0)}{\sin(2\Theta_0)}.
\eee

The goal is to see where this function can be inverted and thus if $\Theta_0\left(\rho\right)$ is a continuous function for any $\theta_h$. Provided $\theta_h\neq n\frac{\pi}{2}$ for $n\in \mathbb{Z}$, we see that $\rho(\Theta_0)\to\infty$  as $\Theta_0\to 0$, and $\rho(\Theta_0)=0$ at $\Theta_0=\theta_h$. We then take the derivative:

\bee
\frac{d\rho}{d\Theta_0}=-\frac{\sin(2\Theta_0)\cos(\theta_h-\Theta_0)+\sin(\theta_h-\Theta_0)2\cos(2\Theta_0)}{\sin^2(2\Theta_0)}
\eee

\begin{figure}
	\centering
	\includegraphics[width=0.6\linewidth]{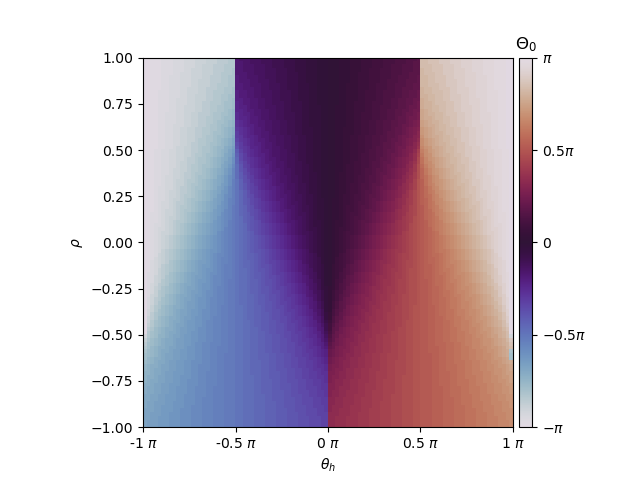}
	\caption{The equilibrium angle $\Theta_0$ of magnetisation as a function of applied magnetic field tilt $\theta_h$ and the ratio of uniaxial anisotropy to Zeeman interaction strength $\rho$. The point $\theta_h=0$, $\rho=-\frac{1}{2}$ corresponds to the second-order phase boundary $h_z+2h_a=0$, and the line $\theta_h=0$, $\rho<-\frac{1}{2}$ corresponds to the symmetry-breaking tilted phase.}
	\label{fig:thetah_rho_phase_diagram}
\end{figure}

If we set this to zero, we find no solution between $0$ and $\theta_h$, meaning that the gradient must always be negative in this region. We can thus invert to give a continuous function $\Theta_0\left(\rho\right)$, bearing in mind that we know $\Theta_0(0)=\theta_h$, $\lim_{\rho\to\infty}\Theta_0\left(\rho\right)=0$. Therefore there are no other first-order transition besides the one already found. The dependence of $\Theta_0(\theta_h,\rho)$ is plotted in Fig. \ref{fig:thetah_rho_phase_diagram}, and we indeed see first-order transitions emerge in $(\theta_h,\rho)$ space at some critical positive and negative $\rho$ for $\theta_h=\frac{\pi}{2}$, $0$ respectively.

It remains to exactly calculate the points in $(\theta_h,\rho)$ space where the two-first-order transitions start. To do this, we consider the function

\bee
\theta_h(\Theta_0)=\Theta_0+\arcsin(\rho\sin(2\Theta_0))
\eee

Note that this function need not be defined for all $\Theta_0$. We then take the derivative:

\bee
\frac{d\theta_h}{d\Theta_0}=1+\rho \frac{2\cos(2\Theta_0)}{\sqrt{1-\rho^2\sin^2(2\Theta_0)}}.
\eee

For positive $\rho$, the minimum value of this gradient is at $\Theta_0=\frac{\pi}{2}$ and is equal to $1-2\rho$. For $\rho<\frac{1}{2}$, this minimum gradient is positive, and thus $\theta_h(\Theta_0)$ is a monotonic function and can be inverted to give a continuous function: there is no first-order transition. Meanwhile, for $\rho>\frac{1}{2}$, the gradient becomes negative and thus the inverse function of $\theta_h(\Theta_0)$ is multivalued for some values of $\theta_h$ around $\frac{\pi}{2}$. By looking for the absolute minimum, we can construct a single-valued function $\Theta_0(\theta_h)$, but it cannot be continuous, given that we know $\Theta_0(0)=0$ and $\Theta_0(\pi)=\pi$. 

The same argument applies for negative $\rho$, where now the minimum value of the gradient is $1+2\rho$ at $\Theta_0=0,\pi$ and when $\rho>-\frac{1}{2}$ the minimum gradient is positive, meaning no first-order phase transition as we vary $\theta_h$. Then for $\rho<-\frac{1}{2}$, there is a first-order transition.

This reproduces exactly the boundaries seen in Fig. \ref{fig:thetah_rho_phase_diagram}. Reintroducing the physical parameters $h_z$ and $h_a$, we get the three-dimensional phase diagram shown in Fig. \ref{fig:3D_phase_diagram} in the main section. 

\section{Decay lengthscale of the general linearised Euler-Lagrange equations \label{appendix_2d_decaylengthscale}}

Let us first decompose $\psi_{1,2}$ into azimuthal Fourier modes:
\bee
\psi_i=\sum_{M=-\infty}^{M=\infty} c_M^i(r)e^{iM\phi},
\eee
where since $\psi_{1,2}$ are real, $c_M^i(r)$ are complex functions obeying $c^i_M(r)=\bar{c}^i_{-M}(r)$.

We can substitute this into the linearised Euler-Lagrange equations \eqref{general_linearised_euler_lagrange} and collect terms with the same factor of $e^{iM\phi}$ to get an infinite series of differential equation relations between the $c_M^i(r)$. 

Defining $a=a_1+ia_2$, the linearised Euler-Lagrange equations then become 
\bee
\begin{pmatrix}
\ldots && 0 && 0 && a\partial_r && 0 \\
0 && -\partial_r^2+\mu_2^2  && -a\partial_r && 0 && 0\\
0 && \bar{a}\partial_r && -\partial_r^2+\mu_1^2 && 0 && 0\\
-\bar{a}\partial_r && 0 && 0 && -\partial_r^2+\mu_2^2 && -a\partial_r\\
0 && 0 && 0 && \bar{a}\partial_r && \ldots\\
\end{pmatrix}
\begin{pmatrix}
\ldots\\
c_{-1}^2(r)\\
c_0^1(r)\\
c_0^2(r)\\
\ldots\\
\end{pmatrix}+O\left(\frac{c^i_M}{r}\right)=\begin{pmatrix}
\ldots\\
0\\
0\\
0\\
\ldots\\
\end{pmatrix}.
\eee

We can neglect $O(c^i_M/r)$ terms since we are looking for exponential-like falloff. By neglecting these terms we remove all dependence on $M$, and see that at sufficiently large $r$ there are four different functions $c^1_{M \text{ odd}}(r)$, $c^1_{M \text{ even}}(r)$, $c^2_{M \text{ odd}}(r)$ and $c^2_{M \text{ even}}(r)$. They are coupled in two separate pairs, each with matrix equation

\bee
\begin{pmatrix}
\ldots && 0 && 0 && 0 && 0 \\
0 && -\partial_r^2+\mu_2^2  && -a\partial_r && 0 && 0\\
0 && \bar{a}\partial_r && -\partial_r^2+\mu_1^2 && a\partial_r && 0\\
0 && 0 && -\bar{a}\partial_r && -\partial_r^2+\mu_2^2 && -a\partial_r\\
0 && 0 && 0 && \bar{a}\partial_r && \ldots\\
\end{pmatrix}
\begin{pmatrix}
\ldots\\
c_{M-1}^2(r)\\
c_M^1(r)\\
c_{M+1}^2(r)\\
\ldots\\
\end{pmatrix}+O\left(\frac{c^i_M}{r}\right)=\begin{pmatrix}
\ldots\\
0\\
0\\
0\\
\ldots\\
\end{pmatrix}.
\eee

We can reduce this to a finite-dimensional matrix equation by considering the radial dependence of $\psi_1$ and $\psi_2$ along $\phi=0$ specifically:

\bee
\psi_{1,2}(r,\phi=0)=\sum_{M=-\infty}^{M=\infty} c_M^{1,2}(r).
\eee

These two functions will be coupled by

\bee
\begin{pmatrix}
 -\partial_r^2+\mu_1^2 && a\partial_r +\bar{a}\partial_r \\
 -\bar{a}\partial_r -a\partial_r && -\partial_r^2+\mu_2^2 \\
\end{pmatrix}
\begin{pmatrix}
\psi_1(r,\phi=0)\\
\psi_2(r,\phi=0)
\end{pmatrix}+O\left(\frac{c^i_M}{r}\right)=\begin{pmatrix}
0\\
0
\end{pmatrix}.
\eee

We see that the equations will be solved to leading order by $\psi_1$, $\psi_2$ of the form $e^{\lambda r}$. The form is nearly identical to the domain wall case and we find

\bee
\lambda = \frac{\pm 1}{\sqrt{2}}\sqrt{\mu_1^2+\mu_2^2  -4a_1^2\pm \sqrt{(\mu_1^2+\mu_2^2 -4a_1^2)^2 - 4\mu_1^2\mu_2^2}}.
\eee

By rotational symmetry, we can say that the decay lengthscale in a given direction will be a function of the component of $\va$ in that direction in general. We thus set $\va$ parallel to the x axis to recover the equation \eqref{2d_decay}.

\bibliographystyle{unsrt}
\bibliography{ThesisBib.bib}

\end{document}